\documentstyle[12pt,twoside,fleqn,espcrc1,epsfig,wrapfig]{article}

\newcommand{\AmS}{{\protect\the\textfont2
  A\kern-.1667em\lower.5ex\hbox{M}\kern-.125emS}}

\title{The $^{4}\mathrm{He}(e,e'p)$ Cross Section at Large Missing Energy}

\author{J. J. van Leeuwe,\address{NIKHEF, P.O. Box 41882, 1009 DB Amsterdam,
           The Netherlands}
        H.P. Blok,\hbox{$^{\rm a,}$}\address{Department of Physics
          and Astronomy, Vrije Universiteit Amsterdam,
          de Boelelaan 1081, 1081 HV Amsterdam, The Netherlands}
        J.F.J. van den Brand,\address{Department of Physics,
          University of Wisconsin, Madison,
          Wisconsin 53706 USA}
        H.J. Bulten,\hbox{$^{\rm c}$}
        G.E. Dodge,\hbox{$^{\rm a,b,}$}\footnote{Permanent Address:
          Department of Physics, Old Dominion University,
          Norfolk, Virigina USA}
        R. Ent,\address{Jefferson Laboratory, 12000 Jefferson Avenue,
          Newport News, Virginia, 23606 USA}
        W.H.A. Hesselink,\hbox{$^{\rm a,b}$}, E. Jans\hbox{$^{\rm a}$},
        W. J. Kasdorp,\hbox{$^{\rm a}$}
        J.M. Laget,\address{Service de Physique Nucl\'eaire,
          Centre d'Etudes de Saclay, 91191 Gif-sur-Yvette Cedex, France}
        L. Lapik\'as,\hbox{$^{\rm a}$}
        C.J.G. Onderwater,\hbox{$^{\rm a,b}$}
        A.R. Pellegrino,\hbox{$^{\rm a,b}$}
        C.M. Spaltro,\hbox{$^{\rm a,b}$}
        J. J. M. Steijger,\hbox{$^{\rm a}$}
        J.A. Templon,\hbox{$^{\rm a,b,}$}\footnote{Permanent
          Address: Dept. of Physics and Astronomy,
          University of Georgia, Athens, Georgia, USA}\footnote{
          Talk given at the 15th International Conference on Few-Body
          Problems in Physics, Groningen, The Netherlands,
          22--26 July 1997}
        O. Unal\hbox{$^{\rm c}$}}

\begin{document}
\maketitle

\begin{abstract}
The $(e,e'p)$ reaction on $^{4}\mbox{He}$ nuclei was studied
in kinematics designed to
emphasize effects of nuclear short-range correlations.
The measured cross sections display a peak in the
kinematical regions where two-nucleon processes are expected
to dominate.
Theoretical models incorporating
short-range correlation effects agree reasonably with the data.
\end{abstract}

%----------------------------------------------------------------------------
% Introduction
%----------------------------------------------------------------------------
\section{Introduction}
\label{sec:intro}

The Independent-Particle Shell Model (IPSM) of nuclei has been
very successful
in providing understanding of many of the observed properties of
nuclei in the ground state and low-lying excited states.
Over the last decade, the IPSM has been
extensively tested by the results of $(e,e'p)$ experiments, which
probe the Fourier transform of individual
nucleon wavefunctions \cite{lapi93,kell96}.
The data indicate that the shell model provides good understanding
of the nucleon wavefunctions, but that states that in the IPSM
are fully occupied, can be depleted by as much as 35\%.
Recent theoretical studies also seem to have converged on such a
picture \cite{pand97}.
The depletion comes about since nucleons are
undergoing violent pairwise collisions which cause a portion
of their wavefunction to be quite different than that predicted by the
IPSM.

These $(e,e'p)$ experiments first concentrated on measurements
in kinematics where the residual $A-1$ nuclear systems
had low recoil momenta and excitation energies, since this is where
the IPSM predicts nucleon strength to reside.
The remaining strength is theoretically
predicted to be found at energies(momenta) above the
Fermi energy(momentum).
Signatures of this strength can be found by performing
$(e,e'p)$ experiments that probe the regions of
large $p_m$ and $E_m$.
This paper describes a measurement of the
$^{4}{\mathrm He}(e,e'p)$
reaction over a wide range of $(E_m,p_m)$
aimed at investigating
the effects of short-range correlations as well as other
two-nucleon processes originating from meson-exchange currents
(MEC) and $\Delta$-isobar currents (IC).

\section{Kinematics}

In an $(e,e'p)$ experiment, the incoming and outgoing electron four-vectors
$k_i$ and $k_f$ determine the energy and momentum transferred to the
nuclear target:
\begin{equation}
  \label{eq:ee}
   q  =  k_i - k_f = (\omega, \vec{q}) .
 \end{equation}
If final-state interactions and many-body currents are ignored,
we can assume that all of the momentum $\vec{q}$ is transferred
to a single nucleon.
Then if this nucleon is detected with momentum $\vec{p}_f$,
its original momentum was
$\vec{p}_i = \vec{p}_f - \vec{q} = -\vec{p}_m$.

If a pair of nucleons is undergoing a strong, short-range interaction,
we expect the nucleons to have high momenta --- one can think of
the Fourier transform of a short-range pair wavefunction, which
must have large components at high momenta.
Due to angular-momentum constraints, we also
expect the momenta of the two nucleons to be
roughly equal in magnitude and opposite in direction.
Thus if we strike one member of this pair which has momentum
$\vec{p}_i$, the other will have momentum ${-}\vec{p}_i$ and
will require an energy $p_i^2/2M$ to be put on shell.
The missing energy for a two-nucleon knockout reaction, of which
only one nucleon is detected, then reads \cite{kest96,ciof96} as:
\begin{equation}
  \label{eq:empm}
  E_m \sim \frac{A-2}{A-1} \frac{p_m^2}{2M} + E_{\rm thr} ,
\end{equation}
recalling that $|\vec{p}_m| = |\vec{p}_i|$.
Thus Eq.~\ref{eq:empm} determines where in $(E_m,p_m)$ space one
should look to find effects of short-range correlations.
It should be noted that Eq.~\ref{eq:empm} is appropriate for
all processes in which two nucleons are knocked out, hence
the strength from
two-body hadronic currents will also follow
this relation.
Such currents include MEC, or $\Delta$ excitation followed
by $\Delta N \rightarrow NN$.

\section{Experiment}

The experiment was performed at NIKHEF in Amsterdam.
The electron beam was provided by the AmPS ring and had an
energy of 525 MeV.
This beam impinged on a cryogenic high-pressure helium cell.
Scattered electrons were detected in the QDQ magnetic
spectrometer of
the EMIN end station, with central values of the energy and
momentum transfer \((\omega, q) = (215 \mbox{ MeV}, 400 \mbox{ MeV}/c)\).
Ejected protons were detected in the HADRON4 detector.
This is a highly-segmented, large-solid-angle (550 msr) device
based on plastic scintillators.
It had a proton energy acceptance of 65--185 MeV, which was determined
by the Pb shielding (5.2 mm thick) placed in front of the detector
to reduce the counting rate in the frontmost scintillators.
Its angular acceptance is $ 40 \times 40 \deg^2$.
The large acceptances of HADRON4 enabled acquisition of an enormous
amount of data in a relatively short time.
Data was obtained over the range
$0 < E_m < 140$ MeV, $20 < p_m < 690$ MeV/{\em c}
using four angular settings of HADRON4.

\section{Results}

\begin{wrapfigure}[31]{l}{72mm}
  \epsfig{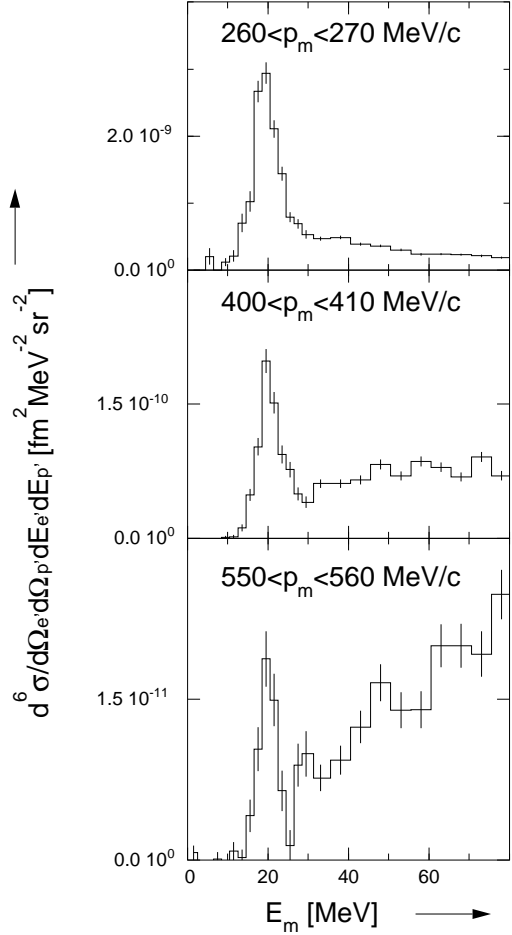}
  \caption{Differential cross sections {\em vs.} \(E_m\) for three
    consecutive bins in \(p_m\).}
  \label{fig:sample}
\end{wrapfigure}

Fig.~\ref{fig:sample} shows sample data from
the experiment.
Differential cross sections
are displayed {\em vs.} $E_m$ for three consecutive bins in $p_m$.
The peak at $E_m \sim 20$ MeV is due to the two-body breakup channel
$^4\mbox{He}(e,e'p)^3\mbox{H}$; an analysis of this portion of
the data has been previously reported \cite{leeu95,leeu97}.
The relative importance of the continuum obviously increases
as $p_m$ increases.
This is in keeping with the observations noted in Sec.~\ref{sec:intro}
that cross sections
for knockout to low-$E_m$ states of the ($A$-1) system
peak at low values of the recoil momentum,
while 
short-distance phenomena will involve relatively larger momenta,
and thus generate strength which peaks at large $E_m$ and $p_m$
in accordance with
Eq.~\ref{eq:empm}.

\begin{wrapfigure}[29]{r}{62mm}
  \epsfig{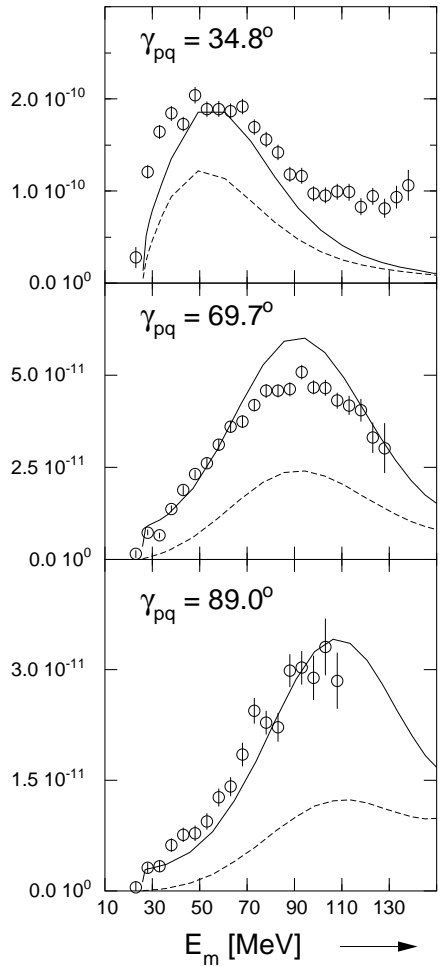}
  \caption{Continuum $(e,e'p)$ differential
    cross sections {\em vs.}\ $E_m$ for three
    values of the
    proton emission angle relative to \(\vec{q}\).
    Dashed: without MEC or IC.
    Solid: full.}
  \label{fig:compare_laget}
\end{wrapfigure}

Fig.~\ref{fig:compare_laget} shows the same data now plotted for
three consecutive bins of
proton emission angle $\gamma_{pq}$ (relative to $\vec{q}$).
The two-body breakup peak has been removed for clarity.
The curves are calculations by Laget.
They include effects from MEC and IC in addition
to the normal one-body hadronic currents.
The $^4{\mathrm He}$ wavefunction was computed with a realistic
$NN$ interaction, so SRC effects are implicitly included in the one-body
hadronic current.
FSI are approximately
accounted for by single-step rescattering diagrams.
The dashed curve shows the results of calculations in which only
one-body currents are taken into account.
They may be considered as a qualitative indication for the importance
of short-range correlations to the differential cross section.

Several features are worth noting in these plots.
The broad bump in the experimental
strength is well-reproduced by the calculation,
and changes with $p_m$ in a way consistent with expectations
from Eq.~\ref{eq:empm}.
This is direct evidence of the dominance of two-nucleon knockout.
At small values of the angle $\gamma_{pq}$ (near parallel kinematics)
the effect of hadronic two-body currents is small and the
cross sections may be largely attributed to the effect of short-range
correlations.
At larger values of $\gamma_{pq}$ the role of the hadronic two-body
currents gradually increases.

Fig.~\ref{fig:compare_laget} also shows that for the lowest
values of $\gamma_{pq}$ a large fraction of the cross section at
the high-$E_m$ side cannot be accounted for by the calculation.
It has been recently suggested \cite{sick96} that this tail is due
to multistep $(e,e'N)(N,p)$ processes,
where the struck nucleon knocks out
one or more additional nucleons as it exits the nucleus.
This conjecture is supported by at least one calculation
\cite{deme97}.

The data have been compared to predictions by Simula
\cite{ciof96,simu96}.
Comparable agreement to what is shown for Laget's calculations
is obtained near parallel kinematics, but the agreement
deteriorates with $\gamma_{pq}$, which is attributed
to the neglect of MECs in Simula's calculations.

\section{Conclusions}

An experiment has been performed studying the $(e,e'p)$ reaction
on $^{4}\mbox{He}$ in kinematics which emphasize
effects related to two-body processes in the nucleus.
The measured cross sections peak at the predicted location,
and reasonably quantitative agreement is achieved with theories
including SRC effects and two-body currents.

\bibliography{mainz}

\end{document}